\documentclass[aps,pre,reprint,onecolumn,superscriptaddress,nobibnotes,nodoi,noeprint]{revtex4-2}

\usepackage{graphicx}
\usepackage{amsmath,amssymb,physics,bm,float}
\usepackage{soul}
\usepackage[colorlinks=true,citecolor=blue,urlcolor=blue,linkcolor=blue]{hyperref}
\usepackage{natbib}
\usepackage{upgreek}
\usepackage{mathtools}
\usepackage{dcolumn}

\newcommand{\Cpp}{C\nolinebreak\hspace{-.05em}\raisebox{.4ex}{\tiny\bf +}\nolinebreak\hspace{-.10em}\raisebox{.4ex}{\tiny\bf +}}

\DeclareMathOperator*{\argmax}{argmax}
\def\rms{\rm\scriptscriptstyle}
\def\dd{{\rm d}}
\def\CC{{C\nolinebreak[4]\hspace{-.05em}\raisebox{.4ex}{\tiny\bf ++}}}

\begin{document}

\title{Fast Brownian cluster dynamics}

\author{Alexander P. Antonov}
\email{alantonov@uos.de}
\affiliation{Universit{\"a}t Osnabr{\"u}ck,
Fachbereich Mathematik/Informatik/Physik,
Institut f{\"u}r Physik,
Barbarastra{\ss}e 7,
D-49076 Osnabr{\"u}ck,
Germany}
\affiliation{Institut f{\"u}r Theoretische Physik II: Weiche Materie, 
Heinrich-Heine-Universit{\"a}t D{\"u}sseldorf, 
Universit\"atsstra{\ss}e 1,
D-40225 D{\"u}sseldorf, 
Germany}

\author{Sören Schweers}
\email{sschweers@uos.de}
\affiliation{Universit{\"a}t Osnabr{\"u}ck, 
Fachbereich Mathematik/Informatik/Physik,
Institut f{\"u}r Physik,
Barbarastra{\ss}e 7,
D-49076 Osnabr{\"u}ck,
Germany}

\author{Artem Ryabov}
\email{rjabov.a@gmail.com}
\affiliation{Charles University, 
Faculty of Mathematics and Physics, 
Department of Macromolecular Physics, 
V Hole\v{s}ovi\v{c}k\'ach 2, 
CZ-18000 Praha 8, 
Czech Republic}

\author{Philipp Maass}
\email{maass@uos.de}
\affiliation{Universit{\"a}t Osnabr{\"u}ck,
Fachbereich Mathematik/Informatik/Physik,
Institut f{\"u}r Physik,
Barbarastra{\ss}e 7,
D-49076 Osnabr{\"u}ck,
Germany}

\date{7 June 2024}

\begin{abstract}
We present an efficient method to perform overdamped Brownian dynamics simulations in external force fields
and for particle interactions that include a hardcore part. The method applies to particle motion in one dimension, 
where it is possible to update particle positions by repositioning particle clusters as a whole. 
These clusters consist of several particles in contact. They form because particle collisions are treated as completely inelastic
rather than elastic ones. Updating of cluster positions in time steps is carried out by cluster fragmentation and merging procedures.
The presented method is particularly powerful at high collision rates in densely crowded systems,
where collective movements of particle assemblies is governing
the dynamics. As an application, we simulate the single-file diffusion of sticky hard spheres in a periodic potential.
\end{abstract}

\maketitle

\section{Introduction}
\label{sec:Introduction}
In many-body dynamics of extended objects, the excluded volume interaction plays a fundamental role.
It takes into account the finite size of particles and
is known also as hardcore or steric interaction. A convenient model for
investigating the impact of excluded volume interactions on the dynamics are systems of hard spheres, which
are widely studied theoretically and experimentally, e.g., in colloidal suspensions \cite{Bartlett/Megen:1994, Royall/etal:2023}.
Even for most softcore interactions, there is a strong repulsive part when particles
approach each other, and this repulsive part can be represented
by a hard-sphere interaction with an effective sphere diameter \cite{Barker/Henderson:1976, Antonov/etal:2021b}.
Many pair interactions can thus be modeled by a hard-sphere repulsion plus additional
contributions accounting for attractive or additional repulsive forces.

While accurate results can be derived for the fundamental hard-sphere interaction 
by advanced analytical methods, it is difficult to deal with it in simulations. This is due to the singular nature
of the interaction force.
The first pioneering simulations of hard sphere systems were carried out
with Monte Carlo simulations by rejecting particle overlaps \cite{Cichocki/Hinsen:1990}.
For Brownian dynamics simulations, many algorithms have been developed.
In one type of algorithms, strong repulsive interaction potentials are used in a thin shell region of the spheres
to limit particle overlaps to a very small region \cite{Heyes/Branka:1994, Heyes/etal:1994, Moriguchi/etal:1995}.
Potential-free algorithms with a repositioning scheme when particles overlap or come in contact were used
in Refs.~\cite{Heyes/Melrose:1993, Schaertl/Sillescu:1994, Foss/Brady:2000, Tao/etal:2006}.
More recent methods use elastic collisions to deal with the hard-sphere interaction \cite{Strating:1999, Miller/Luding:2004, Scala/etal:2007, Scala:2012}. 
The corresponding algorithms 
were shown to yield correct statistical measures for the realizations of the Brownian stochastic process
for spatially constant external forces in the limit of small time steps \cite{Strating:1999, Scala:2012}.
For spatially varying external forces, good results could be obtained by an approximate scheme based
on the known transition probability for a Brownian particle 
on a half-line with reflecting boundary conditions \cite{Behringer/Eichhorn:2012}. In highly crowded systems,
where a large number of collisions occurs 
even in small time intervals, the use of elastic collisions becomes computationally demanding.

Recently, we have introduced the Brownian cluster dynamics (BCD)
method for simulating one-dimensional systems of hard spheres
\cite{Antonov/etal:2022c}, which has a
computational time of order $\mathcal{O}(N^2)$.
Here we present a major improvement that reduces the computational time to $\mathcal{O}(N)$.
It thus allows one to effectively tackle even very high numbers of collisions in densely crowded systems,
where cluster dynamics is
inherent and leads to new physical phenomena such as solitary cluster waves
\cite{Antonov/etal:2022a, Cereceda-Lopez/etal:2023, Antonov:2023, Antonov/etal:2024}.
The BCD method is particularly suitable
for dealing with cooperative particle movements due to propagation,
merging and fragmentation of particle clusters.  It provides a means also for
simulating Brownian dynamics of sticky hard spheres with attractive
contact interactions as described by Baxter's model
\cite{Baxter:1968}, and more generally for arbitrary interactions with
hardcore part. 

The paper is organized as follows. In Sec.~\ref{sec:cluster-dynamics}, we
describe the underlying principles
of the BCD algorithm. It is based on conditions on the total forces acting on the particles that
determine how a cluster fragments into subclusters. 
Section~\ref{sec:fragmentation} shows how fragmentations of clusters can be carried out efficiently in spite of the fact that the number
of possible fragmentations grows exponentially with the cluster size.
In Sec.~\ref{sec:merging}, we discuss an efficient premerging procedure, where one
avoids to perform particle collisions in chronological order. The detailed simulation method
is presented in Sec.~\ref{sec:algorithm} and an application to single-file diffusion of sticky hard spheres in a periodic potential
in Sec.~\ref{sec:application}.

\section{Particle cluster dynamics}
\label{sec:cluster-dynamics}
Overdamped Brownian dynamics of $N$ particles
in one dimension is described by the Langevin
equations
\begin{equation}
\frac{\dd x_i}{\dd t}=\mu f(x_i)+ \mu\hspace{-0.5em}
\sum_{j=1, j\ne i}^N\hspace{-0.3em} f_{\rm int}(x_i,x_j)+\sqrt{2D}\,\xi_i(t)\,,
\hspace{1em}i=1,\ldots,N\,,
\label{eq:langevin}
\end{equation}
where $x_i$ are the particle positions,
$f(x)$ is an external force, $f_{\rm int}(x_i,x_j)$ a pair
interaction force, $\mu$ the particle mobility, $D = k_{\rm B}T\mu$ the diffusion coefficient,
$k_{\rm B} T$ the thermal energy, and
$\xi_i(t)$ are stationary Gaussian stochastic processes with zero mean
and correlation functions
$\langle\xi_i(t)\xi_j(t')\rangle=\delta_{ij}\delta(t-t')$.  
The particle coordinates obey
\begin{equation}
|x_i-x_j|\ge\sigma\,,
\end{equation}
where $\sigma$ is the hard-sphere diameter. This hard-sphere interaction is not
included in $f_{\rm int}(x_i,x_j)$ in Eq.~\eqref{eq:langevin}.  Its
consideration in Brownian dynamics simulations requires a special
treatment.

When treating particle collisions as completely inelastic, clusters of neighboring particles in contact 
can form even in the absence of adhesive interactions.
A cluster of $n$ particles in contact is called an $n$-cluster. Single
particles are 1-clusters.  One can distinguish two types of contacts
in an $n$-cluster: unstable contacts, which disappear due to
separation of particles in the course of their motion, and stable
contacts, which remain.  At each unstable contact there is a
fragmentation into two subclusters.  As there are $(n\!-\!1)$ contacts
in an $n$-cluster, it can decompose into one of $2^{n-1}$ different
fragmentations.  If there are $(r\!-\!1)$ unstable contacts
($r=1,\ldots,n$), $r$ subclusters of sizes $m_1,\ldots,m_r$ form,
$m_1+\ldots+m_r=n$.  The case $r=1$ means that the cluster does
not fragment (no unstable contacts), and the case $r=n$ that the
cluster fragments completely into single particles (all contacts
unstable).

Which fragmentation occurs depends on the total forces
\begin{equation}
F_i=f(x_i)+\hspace{-0.5em}\sum_{j=1, j\ne i}^N\hspace{-0.3em} f_{\rm int}(x_i,x_j)
+\frac{\sqrt{2D}}{\mu}\,\xi_i(t)\,,
\hspace{1em}i=1,\ldots,n\,,
\label{eq:Ftot}
\end{equation}
acting on the particles in the cluster.  For the decomposition
$(m_1,\ldots,m_r)$ to occur, the particles within each subcluster $k$,
$k=1,\ldots,r$, have to fulfill the non-splitting conditions
\begin{subequations}
\begin{equation}
\frac{1}{l}\sum\limits_{j=1}^l F_{m_1+\ldots+m_{k-1}+j}^{\rm tot}
\ge\frac{1}{m_k\!-\!l}\sum\limits_{j=l+1}^{m_k}
F_{m_1+\ldots+m_{k-1}+j}^{\rm tot}\,,\hspace{1em} l=1,\ldots,m_k-1\,,
\label{eq:nonsplitting-subclusters}
\end{equation}
and the non-splitting conditions must by violated for all unstable
contacts,
\begin{equation}
\bar F_k=\frac{1}{m_k}\sum\limits_{j=1}^{m_k}F_{m_1+\ldots+m_{k-1}+j}^{\rm tot}
<\frac{1}{m_{k+1}}\sum\limits_{j=1}^{m_{k+1}} F_{m_1+\ldots+m_k+j}^{\rm tot}=\bar F_{k+1}\,,
\hspace{1em}k=1,\ldots,r-1\,.
\label{eq:splitting-subclusters}
\end{equation}
\end{subequations}
Here, $\bar F_k^{\rm tot}$ is the mean total force acting on
subcluster $k$. Equations~\eqref{eq:nonsplitting-subclusters} and
\eqref{eq:splitting-subclusters} determine how a cluster fragments.
Given a fragmentation, the subclusters $k$ move with velocities
\begin{equation}
v_k=\frac{\bar F_k}{\mu}\,,\hspace{1em} k=1,\ldots,r\,,
\label{eq:vk}
\end{equation}
until they get in contact with another cluster or get further
fragmented due to the change of the total forces.

When two clusters of sizes $n_1$ and $n_2$ with velocities $v_1$ and
$v_2$ get in contact and the non-splitting conditions are fulfilled
for the merged cluster of size $n=n_1+n_2$, the merged cluster moves
with velocity
\begin{equation}
v=\frac{n_1v_1+n_2v_2}{n_1+n_2}=\frac{1}{n}(n_1v_1+n_2v_2)\,.
\label{eq:v-after-merging}
\end{equation}
This corresponds to a completely inelastic collision of the $n_1$- and
$n_2$-clusters.

Equations~\eqref{eq:nonsplitting-subclusters}-\eqref{eq:v-after-merging}
are the basis of the BCD method.  There are various possibilities for
an algorithmic implementation of the method, both with variable time
step and fixed time step.

Here we will use a fixed time step $\Delta t$ and consider the forces
$F_j(t)$ to be constant during each time interval $[t,t+\Delta t[$ as
in the Euler-Maruyama method \cite{Kloeden/Platen:1992, Saito/Mitsui:1993}.
At the beginning of each time
step $t_0+j\Delta t$, $j=0,1,\ldots$,
we perform the fragmentation of all clusters according to Eqs.~\eqref{eq:nonsplitting-subclusters} 
and \eqref{eq:splitting-subclusters}. Because the forces are kept
constant, no fragmentation of clusters occurs during time intervals
$]t_0+j\Delta t,t_0+(j+1)\Delta t[$.  Accordingly, clusters merge only during each time step,
and the merged clusters propagate with
velocities given by Eq.~\eqref{eq:v-after-merging}. The updating
of particle positions thus proceeds in sequential application of
fragmentation and merging. In the next two sections, we discuss
how fragmentation and merging are carried out.

\section{Fragmentation procedure}
\label{sec:fragmentation}
There are $2^{n-1}$ possible fragmentations of an
$n$-cluster. Inequalities~\eqref{eq:nonsplitting-subclusters} and
\eqref{eq:splitting-subclusters} determine which of them occurs.  To
check these inequalities for all possible fragmentations would require
an $\mathcal{O}(2^n)$ computational effort.  Fortunately, this high exponential
effort in the cluster size can be avoided by obtaining the correct fragmentation for an
$n$-cluster from an iterative \textit{pair splitting procedure}, where one
needs to check at most $\mathcal{O}(n^2)$ conditions only.

In the procedure, we consider the $(n\!-\!1)$ possible divisions of an
$n$-cluster into pairs $(j,n\!-\!j)$ of subclusters.  A division
$(j,n\!-\!j)$ is called a pair splitting, if the mean force $\bar
F_j^-$ on the particles in the left subcluster is smaller than the
mean force $\bar F_j^+$ on the particles in the right subcluster,
i.e.\ when
\begin{equation}
\bar F_j^+-\bar F_j^-=\left(\frac{1}{n-j}\sum_{i=j+1}^n
F_i-\frac{1}{j}\sum_{i=1}^j F_i\right)>0.
\label{eq:pair-splitting}
\end{equation}
In such a case, we call $j$ a pair splitting point.  If there is no
pair splitting, the $n$-cluster moves as a whole. If there is only one
pair splitting at point $s$, the $n$-cluster fragments into two
subclusters of size $s$ and $(n\!-\!s)$.

However, if there is more than one pair splitting, a pair splitting
point $j$ does not need to be an unstable contact between particles
$j$ and $j+1$. This is because the full fragmentation of an
$n$-cluster is determined by the conditions
\eqref{eq:nonsplitting-subclusters} and
\eqref{eq:splitting-subclusters}, and not by the conditions for pair
splittings. For example, for a 3-cluster with forces $F_1$, $F_2$, and
$F_3$ satisfying $F_1=F_3/4$ and $F_2=3F_3/2$, $j_1=1$ and $j_2=2$ are
pair splitting points as $F_1<(F_2+F_3)/2$ and $(F_1+F_2)/2<F_3$. The
contact between particle $2$ and $3$, however, is stable because
$F_2>F_3$, i.e.\ $j_2=2$ is not an unstable contact.

\begin{figure}[t!]
\includegraphics[width=0.5\textwidth]{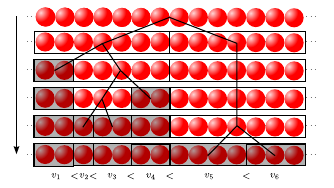}
\caption{Illustration of a fragmentation of a 14-cluster by a sequence
  of five pair fragmentations, leading to six subclusters of sizes
  $m_1=2,\ldots,m_6=3$ with center of mass velocities
  $v_1<\ldots<v_6$. The first pair fragmentation is the principal one.
  It splits the 14-cluster into a pair of subclusters (framed boxes)
  of size $7$ at a point where the difference between mean forces
  acting on two subclusters is largest,
  cf.\ Eqs.~\eqref{eq:pair-splitting},
  \eqref{eq:pair-fragmentation}. The two subclusters then are further
  pair-fragmented at points of largest mean force difference and so on
  until all subclusters have no pair splitting points (shaded
  clusters).  Solid lines connect center of masses of subclusters.}
\label{fig:fragmentation_illustration}
\end{figure}

Intuitively, one can expect the pair splitting point
\begin{equation}
s=\argmax\limits_{1 \le j \le n-1}\{\bar F_j^+ - \bar F_j^-\,|\, (\bar F_j^+ - \bar F_j^-)>0\}
\label{eq:pair-fragmentation}
\end{equation}
with largest force difference $(\bar F_j^+ -\bar F_j^-)>0$ to give an
unstable contact in the fragmentation of an $n$-cluster. 
For non-unique maximum in Eq.~\eqref{eq:pair-fragmentation}, 
one can choose the pair fragmentation with the smallest $s$.  
If the pair splitting point with largest force difference
gives an unstable contact, one can proceed by iteration: for each of the two
subclusters of sizes $s$ and $(n\!-\!s)$ again the pair splittings
with largest force differences are determined, and so on, until there
are no further pair splittings.  This iterative procedure, illustrated
in Fig.~\ref{fig:fragmentation_illustration}, was introduced
heuristically in our previous algorithmic implementation of BCD
\cite{Antonov/etal:2022c} and will be proven here. A pair splitting
point fulfilling Eq.~\eqref{eq:pair-fragmentation} is called a
pair fragmentation point. We want to prove that the full fragmentation
of an $n$-cluster can be done by successive pair fragmentations.

Let $K$ be the total number of successive pair fragmentations of an
$n$-cluster, implying that there are no further pair splittings within
the $K+1$ resulting subclusters $j=1,\ldots,K+1$.  Accordingly, for
each subcluster $j$, the non-splitting conditions
\eqref{eq:nonsplitting-subclusters} are obeyed. To prove that the
fragmentation is correct, the splitting conditions
\eqref{eq:splitting-subclusters} must be satisfied also. Rewriting
them in terms of the velocities \eqref{eq:vk}, these conditions are
$v_1<v_2\ldots<v_{K+1}$.  Hence, we need to show this ordering of
velocities.  In fact, we show that a corresponding ordering of
velocities
\begin{equation}
v_1<v_2\ldots<v_{k+1}
\label{eq:vorder}
\end{equation}
is obtained after any number $k$, $k=1,\ldots, K$ of successive pair
fragmentations.

We perform the proof by complete induction with respect to the number
$k$ of pair fragmentations.  Below we show that the inequalities
\eqref{eq:vorder} hold true for $k=1$, 2, and 3, which form the base
cases of the induction.  Assuming the inequalities \eqref{eq:vorder}
to be fulfilled for $k'\le k$, $k>3$, we need to show that they are
obeyed for $k+1$ also.

\begin{figure}[t!]
\includegraphics[width=1\textwidth]{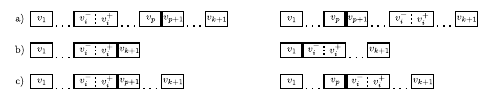}
\caption{Illustration of the three different cases~a)-c) treated in the proof of the pair fragmentation procedure. 
In all cases, the principal pair fragmentation point~$p$ (see text) 
is marked by a thick line. The $(k+1)$th fragmentation at pair fragmentation point~$i$ is marked by a dashed line.
In case~a), the $(k+1)$th pair fragmentation splits a 
subcluster that is neither the $p$th nor the $(p+1)$th subcluster, i.e.\ a subcluster not next to the principal pair fragmentation point~$p$.
In case~b), the $(k+1)$th pair 
fragmentation at point~$i$ splits a subcluster next to $p$ and the other subcluster next to $p$ is the last or first subcluster in the sequence.
In case~c) the situation is the same 
as in case~b) except that now the other subcluster next to $p$ is neither the last nor first subcluster in the sequence.}
\label{fig:fragmentation_proof_cases}
\end{figure}

After $k$ pair fragmentations, the original $n$-cluster is fragmented
into $k+1$ clusters $j=1,\ldots,k+1$ of sizes $m_j$, i.e.\ into an
ordered set $(m_1,\ldots,m_{k+1})$ of clusters, where
$\sum_{j=1}^{k+1}m_j=n$. Let the first pair fragmentation of the
$n$-cluster be between subclusters $p$ and $p+1$, denoted
$(m_1,\ldots,m_p\,|\,m_{p+1},\ldots,m_{k+1})$.  For this principal
pair fragmentation of the $n$-cluster given by
Eq.~\eqref{eq:pair-fragmentation}, the pair splitting point
$s=m_1+\ldots+m_p$ has largest force difference.  A further $(k+1)$th
pair fragmentation will split one of the subclusters $j=1,\ldots,k+1$
into a pair of clusters, say the subcluster $i$ into two clusters of
sizes $m_i^-$ and $m_i^+$, $m_i^-+m_i^+=m_i$. Their
velocities are $v_i^-$ and $v_i^+$, and the velocity $v_i=(m_i^-v_i^-+m_i^+v_i^+)/m_i$ 
of their parent cluster satisfies
$v_i^-<v_i<v_i^+$. We now
consider three cases \newcounter{listnum}
\begin{list}{\alph{listnum})}{\setlength{\leftmargin}{1.5em}\setlength{\rightmargin}{0em}
\setlength{\itemsep}{0ex}\setlength{\topsep}{1ex}\usecounter{listnum}}

\item $i\in\{1,\ldots,p-1\}$ or $i\in\{p+2,\ldots,k+1\}$, 

\item $i=p=k$ or $i=p+1=2$,

\item $i=p<k$ or $i=p+1>2$\,.

\end{list}
The cases are illustrated in Fig.~\ref{fig:fragmentation_proof_cases}.
To show the ordering of velocities
$v_1<\ldots<v_{i-1}<v_i^-<v_i^+<v_{i+1}<\ldots<v_{k+1}$,
we use in case~a) the induction assumption for $k=1$, in case~b) for $k=2$,
and in case~c) for $k=3$. That we need $k=1$, 2, and 3 as base cases for the
induction is due to the fact that the case $k=2$ does not follow
from $k=1$, and the case $k=3$ does not follow from $k=1$ and
$k=2$. However, the case $k=4$ can be treated knowing that the
velocity ordering \eqref{eq:vorder} is valid for $k=1$, 2 and 3.

\vspace{1ex} \textbf{Case a)} If $i\in\{1,\ldots,p-1\}$, the $(k+1)$th
pair fragmentation yields in total $p+1$ subclusters left of the
principal pair fragmentation. Because $p\le k$, there are at most
$k+1$ subclusters to the left.  All these subclusters are resulting
from at most $k$ pair fragmentations of the single left cluster
obtained after the first principal pair fragmentation. According to
the induction assumption, we thus have
$v_1<\ldots<v_i^-<v_i^+<\ldots<v_p$.  Also we have $v_p<v_{p+1}<\ldots<v_{k+1}$, 
because $v_1<\ldots<v_i<\ldots<v_p<v_{p+1}<\ldots<v_{k+1}$ holds for the $k+1$ 
subclusters before the $(k+1)$th pair fragmentation. Hence, after $k+1$ pair
fragmentations, all velocities of the $k+2$ subclusters are ordered,
$v_1<\ldots<v_i^-<v_i^+<\ldots<v_{k+1}$. An analogous reasoning
gives the full ordering of velocities if $i\in\{p+2,\ldots,k+1\}$.

\vspace{1ex} \textbf{Case b)} For $i=p=k$, the $(k+1)$th pair
fragmentation splits the $k$th subcluster into two clusters of sizes
$m_k^-$ and $m_k^+$, with velocities $v_k^-<v_k<v_k^+$. By the
induction assumption we know that $v_1<\ldots<v_k^-<v_k^+$ and it
remains to be shown that $v_k^+<v_{k+1}$. To this end, we combine the
subclusters of sizes $m_1\ldots,m_{k-1},m_k^-$ to one virtual cluster
of size $m_{\rms L}=m_1+\ldots+m_{k-1}+m_k^-$ with velocity
\begin{equation}
v_{\rms L}=\frac{1}{m_{\rms L}}(m_1v_1+\ldots+m_{k-1}v_{k-1}+m_k^-v_k^-)\,.
\end{equation}
This leads to a virtual decomposition $(m_{\rms L},m_k^+\,|\,m_{k+1})$
of the $n$-cluster into three subclusters, where $v_{\rms L}<v_k^+$.

For the whole $n$-cluster, the force difference at the pair splitting
point $m_{\rms L}$ must be smaller than or equal to the force difference at the fragmentation point
$m_{\rms L}+m_k^+$:
\begin{equation}
\frac{m_k^+v_k^++m_{k+1}v_{k+1}}{m_k^++m_{k+1}}-v_{\rms L} 
\le v_{k+1}-\frac{m_{\rms L}v_{\rms L}+m_k^+v_k^+}{m_{\rms L}+m_k^+}\,.
\label{eq:vdifference-k2}
\end{equation}
The weighted averages are equal to center of mass velocities of the
cluster of sizes $m_k^++m_{k+1}$ and $m_{\rms L}+m_k^+$ resulting from
a corresponding pair-splitting of the $n$-cluster.  Because $v_{\rms L}<v_k^+$,
we obtain, when replacing the second term on the
right-hand side of the inequality \eqref{eq:vdifference-k2} by
$v_{\rms L}$, the desired relation 
\begin{equation}
v_k^+<v_{k+1}
\end{equation}
and the full
ordering $v_1<\ldots<v_k^-<v_k^+<v_{k+1}$.

The case $i=p+1=2$ can be treated analogously by combining the
subclusters of sizes $m_2^+,m_3,\ldots,m_{k+1}$ to one virtual cluster
of size $m_{\rms R}=m_2^++m_3+\ldots+m_{k+1}$, which gives us a
virtual decomposition $(m_1|m_2^-,m_{\rms R})$ into three subclusters
with velocities $v_2^-<v_{\rms R}$.  Comparing again velocity
differences at pair splitting points $m_1$ and $m_1+m_2^-$ of the
$n$-cluster, we obtain $v_1<v_2^-$ and the full ordering
$v_1<v_2^-<v_2^+<v_3<\ldots<v_{k+1}$.

\vspace{1ex} \textbf{Case c)} For $i=p<k$, the $(k+1)$th pair
fragmentation splits the $p$th subcluster into two clusters of sizes
$m_p^-$ and $m_p^+$, with velocities $v_p^-<v_p<v_p^+$.  By the
induction assumption we know that $v_1<\ldots<v_p^-<v_p^+$ and
$v_{p+1}<\ldots<v_{k+1}$, and it remains to be shown that
$v_p^+<v_{p+1}$. To this end, we combine the subclusters of sizes
$m_1\ldots,m_{p-1},m_p^-$ and the subclusters of sizes
$m_{p+2},\ldots,m_{k+1}$ to virtual clusters of sizes 
$m_{\rms L}=m_1+\ldots+m_{p-1}+m_p^-$ and 
$m_{\rms R}=m_{p+2}+\ldots+m_{k+1}$.  Their velocities are
\begin{subequations}
\begin{align}
v_{\rms L}&=\frac{1}{m_{\rms L}}(m_1v_1+\ldots+m_{p-1}v_{p-1}+m_p^-v_p^-)\,,\\[1ex] 
v_{\rms R}&=\frac{1}{m_{\rms R}}(m_{p+2}v_{p+2}+\ldots+m_{k+1}v_{k+1})\,.
\end{align}
\end{subequations}
This leads to a virtual decomposition $(m_{\rms
  L},m_p^+\,|\,m_{p+1},m_{\rms R})$ of the $n$-cluster into four
subclusters, where $v_{\rms L}<v_p^+$ and $v_{p+1}<v_{\rms R}$.

For the whole $n$-cluster, the force differences at the pair splitting
points $m_{\rms L}$ and $m_{\rms L}+m_p^++m_{p+1}$ must be smaller
than at the principal pair fragmentation point $m_{\rms L}+m_p^+$:
\begin{subequations}
\label{eq:vdifference-k3}
\begin{align}
\frac{m_p^+v_p^++m_{p+1}v_{p+1}+m_{\rms R}v_{\rms R}}{m_p^+
+m_{p+1}+m_{\rms R}}-v_{\rms L} 
&\le \frac{m_{p+1}v_{p+1}+m_{\rms R}v_{\rms R}}{m_{p+1}+m_{\rms R}}
-\frac{m_{\rms L}v_{\rms L}+m_p^+v_p^+}{m_{\rms L}+m_p^+}\,,
\label{eq:vdifference-k3-a}\\[1ex]
v_{\rms R}-\frac{m_{\rms L}v_{\rms L}+m_p^+v_p^+
+m_{p+1}v_{p+1}}{m_{\rms L}+m_p^++m_{p+1}}
&\le\frac{m_{p+1}v_{p+1}+m_{\rms R}v_{\rms R}}{m_{p+1}+m_{\rms R}}
-\frac{m_{\rms L}v_{\rms L}+m_p^+v_p^+}{m_{\rms L}+m_p^+}\,.
\label{eq:vdifference-k3-b}
\end{align}
\end{subequations}
These inequalities can be rearranged into an equivalent form
containing the velocity differences between neighboring of the four
clusters:
\begin{subequations}
\label{eq:vdifference-k3-2}
\begin{align}
v_{p+1}-v_p^+&\ge \frac{m_{\rms L}+m_p^++m_{p+1}}{m_{p+1}+
m_{\rms R}}(v_{\rms R}-v_{p+1})- \frac{m_{\rms L}}{m_{\rms L}+m_p^+}(v_p^+-v_{\rms L})\,,
\label{eq:vdifference-k3-2-a}\\[1ex]
v_{p+1}-v_p^+&\ge\frac{m_p^++m_{p+1}+m_{\rms R}}{m_{\rms L}+
m_p^+}(v_p^+-v_{\rms L})- \frac{m_{\rms R}}{m_{p+1}+m_{\rms R}}(v_{\rms R}-v_{p+1})\,.
\label{eq:vdifference-k3-2-b}
\end{align}
\end{subequations}
The velocity differences $(v_p^+-v_{\rms L})$ and 
$(v_{\rms R}-v_{p+1})$ on the right-hand sides of \eqref{eq:vdifference-k3-2}
are positive. If either of the expressions on the right-hand sides of
\eqref{eq:vdifference-k3-2} would be greater than zero, we would
obtain the desired inequality $(v_{p+1}-v_p^+)>0$.

If the right-hand side of \eqref{eq:vdifference-k3-2-a} is greater
than zero, $(v_{p+1}-v_p^+)>0$.  If it is not positive,
\begin{equation}
v_p^+-v_{\rms L}\ge\frac{(m_{\rms L}+m_p^+)(m_{\rms
    L}+m_p^++m_{p+1})}{m_{\rms L}(m_{p+1}+m_{\rms R})}(v_{\rms
  R}-v_{p+1})\,.
\label{eq:vp_vL>}
\end{equation}
We then replace $(v_p^+-v_{\rms L})$ in
inequality~\eqref{eq:vdifference-k3-2-b} by the right-hand side of
\eqref{eq:vp_vL>}, yielding
\begin{align}
v_{p+1}-v_p^+&\ge \frac{v_{\rms R}-v_{p+1}}{m_{\rms L}(m_{p+1}+m_{\rms
    R})} (m_p^++m_{p+1})(m_{\rms L}+m_p^++m_{p+1}+m_{\rms R})>0\,.
\end{align}
This completes the proof that
$v_1<\ldots<v_p^-<v_p^+<v_{p+1}<\ldots<v_{k+1}$\,.

The case $i=p+1>2$ can be treated analogously. It holds
$v_1<\ldots<v_p$ and $v_{p+1}^-<v_{p+1}^+<v_{p+2}<\ldots<v_{k+1}$ due
to the induction assumption, and the subclusters of sizes
$m_1\ldots,m_{p-1}$ as well as the subclusters of sizes
$m_{p+1}^+,m_{p+2},\ldots,m_{k+1}$ are combined to virtual clusters of
sizes $m_{\rms L}$ and $m_{\rms R}$, respectively. This leads to a
virtual decomposition $(m_{\rms L},m_p\,|\,m_{p+1}^-,m_{\rms R})$ of
the $n$-cluster into four subclusters, where $v_{\rms L}<v_p$ and
$v_{p+1}^-<v_{\rms R}$. The reasoning to obtain $v_p< v_{p+1}^-$ for
this virtual decomposition then is carried out in the same manner as
above.

\vspace{1ex} \textbf{Base cases of induction}. We need to prove the
ordering \eqref{eq:vorder} of velocities for $k=1$, 2 and
3. For $k=1$, $v_1^-<v_1^+$ by definition
[Eqs.~\eqref{eq:pair-splitting}, \eqref{eq:pair-fragmentation}].  For
$k=2$, we can follow the reasoning in case b) above without making 
use of any induction assumption. The only difference
is that no combination of subclusters left or right of the principal
pair fragmentation is necessary, i.e.\ $m_{\rms L}=m_1^-$ for $i=1$,
corresponding to the decomposition $(m_1^-,m_1^+\,|\,m_2)$, and
$m_{\rms R}=m_2^+$ for $i=2$, corresponding to
$(m_1\,|\,m_2^-,m_2^+)$. For $k=3$, the validity of \eqref{eq:vorder}
follows for $i\ne p,p+1$ by induction as described above in case a),
for $i=2=p$ or $i=2=p+1$ from the reasoning in case b), and for
$i=1=p$ or $i=3=p+1$ from the reasoning in case c).

\section{Premerging procedure}
\label{sec:merging}
At time $t+\Delta t$, before starting a new fragmentation, the
particle configuration consists of clusters with certain sizes at
certain positions. We define a cluster's position as that of its
center of mass (CM).  Each of these clusters was either already
present or resulted from a merging of a certain number clusters
present after the fragmentation performed at time $t$.

Let us consider an $n$-cluster before fragmentation at time $t+\Delta
t$, which resulted from $k$ binary mergings (inelastic collisions) of
clusters of sizes $n_i$ present at time $t$, $n=\sum_{i=1}^kn_i$. In
the course of the particles' motion within the time interval
$]t,t+\Delta t[$, the mergers take place in a particular order, which
is determined by the initial positions and velocities $x_i$ and
$v_i$ of the clusters $i=1,\ldots,k$ at time $t$.  Simulating 
the cluster trajectories requires a high computational effort
for large $k$.

\begin{figure}[t]
\includegraphics{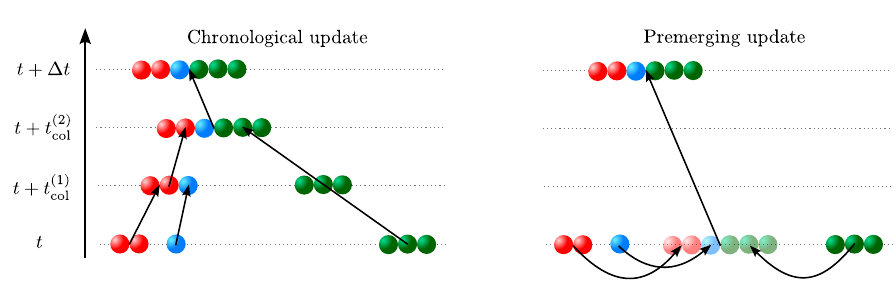}
\caption{Illustration of the chronological and the premerging update
  for three clusters of sizes $n_1=2$ (red), $n_2=1$ (blue), and
  $n_3=3$ (green), having CM positions $x_1$, $x_2$, $x_3$
  and velocities $v_1$, $v_2$, $v_3$ at time $t$. In the chronological update, the
  three clusters move until time $t+t_{\rm col}^{(1)}$, where the 2-
  and 1-cluster get merged in an inelastic collision. Afterwards, the
  resulting merged cluster of size $n_1+n_2=3$ moves with velocity
  $v=(n_1v_1+n_2v_2)/(n_1+n_2)$.  At the later time $t+t_{\rm
    col}^{(2)}$, this 3-cluster collides with the green 3-cluster
  originally at position $x_3$, leading to a merged 6-cluster, which
  moves with velocity $v_{\rms CM}=[(n_1+n_2)v+n_3v_3]/(n_1+n_2+n_3)$.
  In the premerging update, the three clusters are premerged at time
  $t$ (indicated by the curved arrows), and the merged cluster of size
  $n=n_1+n_2+n_3=6$ is placed at $x_{\rms
    CM}=(n_1x_1+n_2x_2+n_3x_3)/n$ and assigned the velocity $v_{\rms
    CM}$.  The straight lines represent uniform linear motions of the
  clusters.  The particle positions (and velocities) at time $t+\Delta
  t$ are the same for both updates.}
\label{fig:premerging_update}
\end{figure}

This can be avoided by taking advantage of the momentum conservation,
which holds true because forces are kept constant in each time
step. The CM of the $k$ clusters moves with its velocity irrespective
of any mergers. Hence, the position of the merged $n$-cluster at time
$t+\Delta t$ can be readily calculated from the CM position and
velocity of the $k$ clusters at time $t$.  We can merge the $k$
clusters already at time $t$ and place the merged $n$-cluster at the
CM position
\begin{equation}
x_{\rms CM}=\frac{1}{n}\sum_{i=1}^k n_ix_i\,,
\end{equation}
 and assign to it the corresponding CM velocity
\begin{equation}
v_{\rms CM}=\frac{1}{n}\sum_{i=1}^k n_iv_i\,.
\end{equation}
The position $x_{\rms CM}$ of the $n$-cluster is then updated to
$x_{\rms CM}+v_{\rms CM}\Delta t$ at time $t+\Delta t$.  An example of
this \textit{premerging update procedure} is given in
Fig.~\ref{fig:premerging_update}.

The premerging update can be performed for any group of neighboring
clusters present at time $t$, provided we know that those clusters
will become merged at time $t+\Delta t$.  However, without simulating
cluster trajectories, it is not clear at a first glance how to
identify those clusters at time $t$ that get merged to single
clusters.  As we show now, there exists an efficient solution for this
identification problem.

The identification of clusters becoming merged can be done based on
binary mergings of neighboring clusters.  Consider two neighboring
clusters of sizes $n_1$, $n_2$ at CM positions $x_1$, $x_2>x_1$ with
velocities $v_1$, $v_2$ at time $t$.  Assuming the two clusters not colliding and
moving independently from all other clusters, their CM positions at
time $t'$ are $x_i(t')=x_i+v_i (t'-t)$, $i=1,2$.  If $v_1>v_2$, they
collide at a time instant when $x_2(t+t_{\rm col})-x_1(t+t_{\rm
  col})=(n_1+n_2)\sigma/2$, giving
\begin{equation}
t_{\rm col}=\frac{(x_2-x_1)-(n_1+n_2)\sigma/2}{v_1-v_2}
\label{eq:t_coll}
\end{equation}
for the collision time. For $t_{\rm col}\le\Delta t$, the collision
occurs in $]t,t+\Delta t]$. Since the collision is inelastic, it leads
to a merger of the two clusters.

The key point now is that possible collisions with other clusters
cannot prevent the two clusters from merging. They can only speed it
up.  To understand this, we first note that other clusters can collide
with the first (second) of the two clusters only if they attach to it
from the left (right) side.  After a collision with the first cluster
from the left, the generated merged cluster has a velocity $v_1'>v_1$,
which follows from momentum conservation.  Likewise, a collision with
the second cluster from the right yields a merged cluster with
velocity $v_2'<v_2$. Hence, irrespective of any number of collisions
from the left and right, the first and second cluster merge in the
interval $]t,t+\Delta t]$, meaning that the rightmost particle of the
first cluster and the leftmost particle of the second cluster come
into contact.

To summarize, any merging between neighboring clusters in 
$]t,t+\Delta t]$, which is identified under neglect of motions of other clusters,
will occur in the full many-body dynamics. We call mergings
between neighboring clusters without consideration of other
clusters at time $t$, binary premergings.  In each binary
premerging, the originally separated clusters are merged, the
resulting merged cluster is placed at the CM position of the two
originally separated clusters, and its velocity is set to the CM
velocity of the two clusters.

To identify \textit{all} mergings in the full many-body dynamics,
binary premergings can be applied iteratively, until there are no
further ones. This is because an equivalent cluster configuration is
obtained after each binary premerging, in the sense that the many-body
dynamics would lead to the same particle configuration at time
$t+\Delta t$ as in the chronological update. Thus, after completing
the iteration, an equivalent initial particle configuration is present
at time $t$, which cannot give rise to any binary collision in the
interval $[t,t+\Delta t]$. All clusters in this initial configuration
move independently in $[t,t+\Delta t]$.

\section{Simulation Algorithm}
\label{sec:algorithm}
The fragmentation and merging procedure for BCD can be implemented in
various ways. We present here an efficient algorithm, which uses a
particle-based labeling of states.

\begin{figure}[t!]
\includegraphics[width=0.4\textwidth]{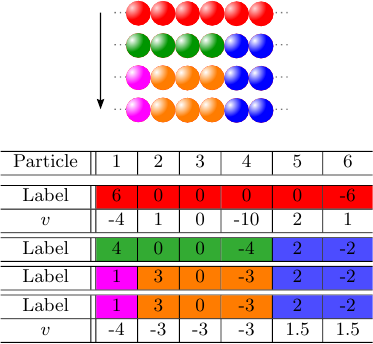}
\caption{
Example illustrating the fragmentation procedure for a 6-cluster. 
The table in the lower part of the figure shows the updating of particle labels and velocities.
The top part shows the initial cluster configuration as well as three further cluster configurations. These configurations
refer to four updating steps given in the partitioning of the table. The updating proceeds from top to bottom, see text. 
Positions and velocities are in some arbitrary units.}
\label{fig:fragmentation_example}
\end{figure}

\begin{figure}[b!]
\includegraphics[width=0.9\textwidth]{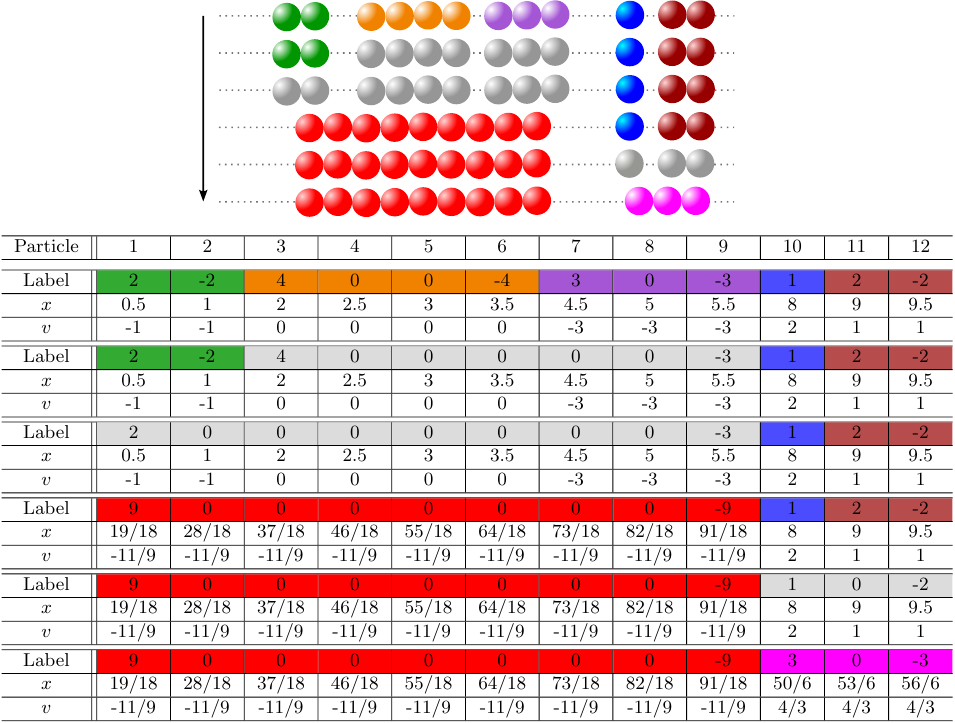}
\caption{Example of premerging procedure for a sequence of 2-, 4-, 3-, 1- and 2-clusters. 
Updatings of particle labels, positions and velocities are listed in the table in the lower part of the figure. In the upper part,
the cluster configuration of the initial state is shown, as well as cluster configurations 
obtained after five updating steps corresponding to the partitioning of the table. 
The updating proceeds from top to bottom, see text. Positions and velocities are in arbitrary units. 
In these units, the particles have a
hard core of size~$\sigma=1/2$ and the time step is $\Delta t=5$.}
\label{fig:merging_example}
\end{figure}

A \CC \ implementation of this algorithm, including
vector manipulation procedures from Refs.~\cite{Sanderson/Curtin:2016, Sanderson/Curtin:2018},
is provided on GitHub~\cite{Antonov/Schweers:2022}. It contains also a treatment of
adhesive contact interactions as described by Baxter's model of sticky
hard spheres \cite{Baxter:1968}. Versions adapted for parallel computing on CPUs and a CUDA-based
GPU implementation are provided as well. Illustrations of the implemented
procedures are given in Figs.~\ref{fig:fragmentation_example} and
\ref{fig:merging_example}.

Overall the algorithm is as follows: At time $t$, the total forces $F_j(t)$ 
acting on the particles 
are calculated from Eq.~\eqref{eq:Ftot}. Then the fragmentation procedure is performed, 
identifying particles forming clusters that do not split at time $t$. The particles in these clusters
stay together during the time step $t\to t+\Delta t$. Further contacts between particles can form
due to cluster merging, which is taken into account by the premerging procedure. After premerging, all particle positions $x_j(t)$ 
are updated to their new positions $x_j(t+\Delta t)$.

In the fragmentation procedure, the initial clusters present at time $t$, resulting from the previous time step, are analyzed
by looping over the clusters from left to right. Given an $n$-cluster, the pair splitting condition \eqref{eq:pair-splitting} is checked
for all $n-1$ possible pair splitting points $j=1,\ldots,n-1$. If there is no pair splitting point, the next cluster to the right is analyzed.
If there are pair splitting points and the largest force difference occurs at position $s$, see Eq.~\eqref{eq:pair-fragmentation}, 
a pair fragmentation is performed at position $s$. In case the maximum is not unique, 
the leftmost position having maximal force difference is chosen for $s$.
The left subcluster of the pair fragmentation is thereafter pair-fragmented. This could lead to the generation of a further left subcluster to be
pair-fragmented. Whenever no pair splitting occurs anymore in the leftmost subcluster, the subcluster (or initial cluster)
to the right is pair-fragmented. 
Hence, in the overall procedure new clusters can emerge, namely as subcluster to the right in the pair splittings, 
but due to the looping over all clusters including the newly formed ones, all initial clusters are considered for fragmentation.
Once the fragmentation procedure has passed through the entire system, all  initial clusters present at time $t$
are fully fragmented. 

To specify which particle belongs to which cluster, we introduce an array of labels, one per particle. 
If a particle is not part of a cluster, its label is set to 1. If it is the leftmost or rightmost particle of an $n$-cluster, its label is set to $n$ or $-n$. 
All remaining labels are set to 0. The encoding of the cluster sizes facilitates the premerging procedure following the fragmentation, see below.

The fragmentation procedure is illustrated in Fig.~\ref{fig:fragmentation_example} for a 6-cluster. 
The table in the lower part of the figure shows the updating of particle labels and particle velocities.
It has four partitions, where the first partition from the top corresponds to an updating of labels and velocities, the second and third division to
an updating of labels, and the fourth to an updating of velocities.
The labels in the first partition are those for the initial 6-cluster.
Total forces calculated for the particles, including noise terms, see Eq.~\eqref{eq:Ftot}, 
give the velocities in the first partition. Between a 4-subcluster to the left and a 2-subcluster to the right, 
a pair splitting has largest force (or velocity) difference: the mean velocity of the 4-cluster is $\bar v_{1234}=(-4+1+0-10)/4=-3.25$ and the 
mean velocity of the 2-cluster is $\bar v_{56}=(2+1)/2=1.5$, yielding $\bar v_{56}-\bar v_{1234}=4.75$.
Accordingly, the 6-cluster is split into a left 4-cluster and a right 2-cluster, yielding the updated labels in the 
second partition of the table. Now the left 4-cluster is checked for the pair fragmentation with largest velocity difference. 
It occurs between a left 1-cluster (single particle) with $v_1=-4$ and a right 3-cluster with $\bar v_{234}=(1+0-10)/3=-3$, giving
$\bar v_{234}-v_1=1$. Thus the 4-cluster is split into a 1- and 3-cluster, giving the updated labels in the 
third partition of the table. The resulting left 1-subcluster cannot fragment, i.e.\ next the neighboring 3-subcluster 
to the right is analyzed. As it has no pair splitting point, the next neighboring 2-subcluster to the right is considered. 
Since this has no pair-splitting point either, the fragmentation procedure is completed by updating the particle 
velocities in the last step: the velocity of the single particle is $v_1$, 
the particles in the 3-cluster get velocities $\bar v_{234}=(1+0-10)/3=-3$, and those in the 2-cluster $\bar v_{56}=1.5$.
These updated velocities are listed in the fourth partition of the table.

The top part in Fig.~\ref{fig:fragmentation_example} illustrates the cluster configurations 
corresponding to the four steps in the fragmentation procedure:
The first row shows the initial 6-cluster (red), the second row
the 4-cluster (green) and 2-cluster (blue) after the first pair fragmentation, and
the third row the 1-cluster (pink), 3-cluster (orange) and 2-cluster (blue) after the second pair fragmentation.
In the last row, the coloring of particles remains the same as only particle velocities are updated.

In the premerging procedure, we loop over the clusters obtained after fragmentation. Each step in the procedure starts by
considering a cluster A and its neighboring one B to the right. Taking the velocities, sizes and positions of the two clusters, 
their possible collision time $t_{\rm col}$ is calculated from Eq.~\eqref{eq:t_coll}. 
If  $t_{\rm col}$ is larger than the time step $\Delta t$, we check for a collision between the right cluster B and the next one C to the right.
If $t_{\rm col}\le\Delta t$, we premerge the two clusters A and B and check for a possible collision between the new premerged AB cluster
and the next neighboring cluster L$_1$ to the left. If $t_{\rm col}\le\Delta t$ for this collision, we premerge the AB cluster with its neighboring cluster L$_1$,
giving an L$_1$AB cluster.
This premerging of clusters to the left is continued until no further collision of a 
corresponding pair of clusters occurs in $\Delta t$, yielding a premerged cluster
L$_k\ldots$L$_1$AB. Thereafter we return to the permerging to the right by checking 
for a collision of the premerged  L$_k\ldots$L$_1$AB cluster with the
C cluster.  This procedure is continued until all neighboring clusters in the system have been checked for merging.

The premerging procedure is illustrated in Fig.~\ref{fig:merging_example} for a sequence of a 2-, 4- 3- 1- and 2-cluster.
Again we show cluster configurations in the upper part of the figure
corresponding to updating steps listed in the table in the lower part. Partitions of the table refer to five steps, where either only certain
particle labels are updated
or all particle labels, positions and velocities of merged clusters. 
In the first partition from the top, we check for a possible collision between the 2-cluster (green in upper part of figure) and the 4-cluster (orange).
Since they move apart ($\bar v_{12}=-1$, $\bar v_{3456}=0$), they cannot collide and we check for a collision between
the next pair of clusters to the right, i.e.\ the 4-cluster and the 3-cluster. 
Calculation of $t_{\rm col}$ for these clusters gives
$t_{\rm col}=1/6\le\Delta t=5$, i.e.\ they merge. We do not immediately update particle positions and velocities according to this merging
but set to zero only labels of particles that become inner particles by the merging,
i.e.\ which are no longer the first or last particle of a cluster.
For the merging of the 4- and 3-cluster, this means that the label -4 of the last particle of the 4-cluster (particle~6 in the table) 
and the label 3 of the first particle of the 3-cluster (particle~7 in the table) are set to zero. The respective two particles are now inner particles of a cluster,
i.e.\ the information of the merging of 4- and 3-cluster is encoded.
Corresponding clusters, which are known to get merged but for which particles positions and velocities are not yet updated, 
are marked in gray in the upper part of the figure. 

We next check whether the 7-cluster obtained from a merger of the 4- and 3-cluster 
would collide with the 2-cluster to the left in time $\Delta t$. As this is the case, the last and first particle of the 2- and 4-cluster
become inner particles of a 9-cluster and the corresponding labels -2 and 4 are set to zero.
No further collision of the 9-cluster is possible with a cluster to the left. It is then checked whether it can collide with the 
next neighboring 1-cluster (blue) to the right. As the 9- and 1-cluster move apart, they are not merged and we
update all particle labels, positions and velocities of the 9-cluster.
Next, it is checked whether the 1-cluster to the right of the 9-cluster and its right neighboring 2-cluster are colliding in $\Delta t$.
This is the case and accordingly the label 2 of particle 11 is set to zero because it becomes an inner particle of the 3-cluster after merging. 
The 3-cluster does not collide with the neighboring 9-cluster to the left and no further collision of it is possible with a cluster to the right.
Hence, all particle labels, positions and velocities of the 3-cluster are updated . This completes the premerging procedure
for this example.

\begin{figure}[b!]
\includegraphics{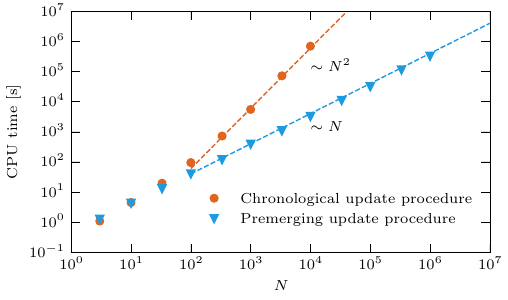}
\caption{ 
Comparison of CPU times when performing simulations of hard spheres in the sinusoidal potential \eqref{eq:U(x)}
on a i5-7600 CPU under periodic boundary conditions with the chronological and 
premerging updates. Parameters are $\sigma=0.99\lambda$,  $U_0=6k_{\rm B}T$, $D/\mu=k_{\rm B}T$, 
and particle number density $\rho=1/\lambda$.  For solving the Langevin equation with the Euler-Maruyama scheme,
the time step $\Delta t=10^{-3}\lambda^2/D$ was chosen. 
The total simulation time is $t_{\rm sim}=3\times10^3\lambda^2/D=3\times 10^6\Delta t$.
The dashed lines indicate the scalings $\sim N^2$ and $\sim N$ of the chronological and premerging updates.}
\label{fig:speed-up}
\end{figure}

When implementing the chronological update procedure, the computational demand increases
as $\mathcal{O}(N^2)$.
This is because the number of collisions per time step is linear in $N$ and for each collision all $N$ particle positions are updated.
By applying the premerging procedure the computational demand is reduced to $\mathcal{O}(N)$.

To demonstrate this improvement, we carried out simulations for both the chronological
and the premerging updates
for a system of hard spheres
in an external sinusoidal potential 
\begin{equation}
U(x)=\frac{U_0}{2}\left[1-\cos(\frac{2\pi x}{\lambda})\right]
\label{eq:U(x)}
\end{equation}
under periodic boundary conditions. 
CPU times obtained from the simulations on a i5-7600 CPU are
shown in Fig.~\ref{fig:speed-up} as a function of $N$. The results in the double-logarithmic
plot confirm the scalings $\sim N^2$ and $\sim N$ for the two methods.

\section{Application to single-file diffusion of sticky hard spheres\\ in periodic potential}
\label{sec:application}
We demonstrate the BCD for single-file diffusion of interacting
particles in a periodic potential.  Generally, Brownian single-file
diffusion occurs in many biological, chemical and engineering systems,
when particle motion is confined to narrow pores \cite{Harris:1965,
  Levitt:1973, Arratia:1983, Hahn/Kaerger:1996, Hahn/etal:1996,
  Hahn/Kaerger:1998, Wei/etal:2000, Cui/etal:2002, Lin/etal:2002,
  Kollmann:2003, Lutz/etal:2004a, Lutz/etal:2004b, Lin/etal:2005,
  Koeppl/etal:2006, Bauer/Nadler:2006, Cheng/Bowers:2007,
  Lizana/Ambjornsson:2008, Kahms/etal:2009, Coste/etal:2010,
  Das/etal:2010, Henseler/etal:2010, Yang/etal:2010, Delfau/etal:2011,
  Dvoyashkin/etal:2013, Dvoyashkin/etal:2014, Kaerger:2014,
  Ryabov:2016, Nygard:2017, Taloni/etal:2017, Chmelik/etal:2018,
  Cao/etal:2018, Zeng/etal:2018, Luan/Zhou:2018, Zhao/etal:2018,
  Dolai/etal:2020, Bukowski/etal:2021, Kaerger/etal:2021,
  Wittmann/etal:2021, Banerjee/etal:2022, Krajnik/etal:2024,
  Sorkin/Dean:2024}.
In applications, it is often important to take into account
inhomogeneous \cite{Barkai/Silbey:2009, Ryabov/Chvosta:2011,
  Sorkin/Dean:2023} or periodic environments
\cite{Taloni/Marchesoni:2006, Dessup/etal:2018, Lips/etal:2018,
  Lips/etal:2019, Ryabov/etal:2019} as well as effects of adhesive
interactions between the particles \cite{Rosenbaum/etal:1996,
  Miller/Frenkel:2004_1, Zaccarelli:2007, Schwarz-Linek/etal:2012,
  Richard/etal:2018, Wang/Swan:2019, Piazza:2014, Bergenholtz:2018,
  Genix/Oberdisse:2018, Assenza/Mezzenga:2019, vonBuelow/etal:2019,
  dePirey/etal:2019, Wang/etal:2019, Whitaker/etal:2019,
  Smith/etal:2020, Bakhshandeh/etal:2020, Balazs/etal:2020}.  Adhesive
interactions lead to pronounced particle clustering at high particle
densities, where fragmentation and merging processes are generally
difficult to deal with in computer simulations
\cite{Bou-Rabee/Holmes-Cerfon:2020, Holmes-Cerfon:2020}. Our BCD
algorithm provides a convenient and efficient method to tackle
single-file dynamics in the presence of pronounced clustering for
arbitrary particle interactions.

As an example, we consider single-file diffusion of sticky hard
spheres \cite{Baxter:1968, Percus:1982} in the potential \eqref{eq:U(x)}.
The pair interaction is
\begin{equation}
\exp[-V(r)/k_{\rms B} T] = \Theta(r-\sigma)+\gamma\lambda\,\delta_+(r-\sigma)\,,
\label{eq:pair-interaction}
\end{equation}
where $\Theta(.)$
is the Heaviside step function [$\Theta(x)=1$ for $x>0$ and zero otherwise]. 
It takes into account the
hardcore repulsion, since $V(r)=\infty$ for $r<\sigma$.
The function $\delta_+(r)$ is the right-sided 
$\delta$-function: for any test function $h(r)$ and $\epsilon>0$, 
it holds $\int_0^\epsilon\dd r\, h(r)\delta_+(r)=h(0)$. The parameter $\gamma$ is the strength of the stickiness.

\begin{figure}[b!]
\includegraphics{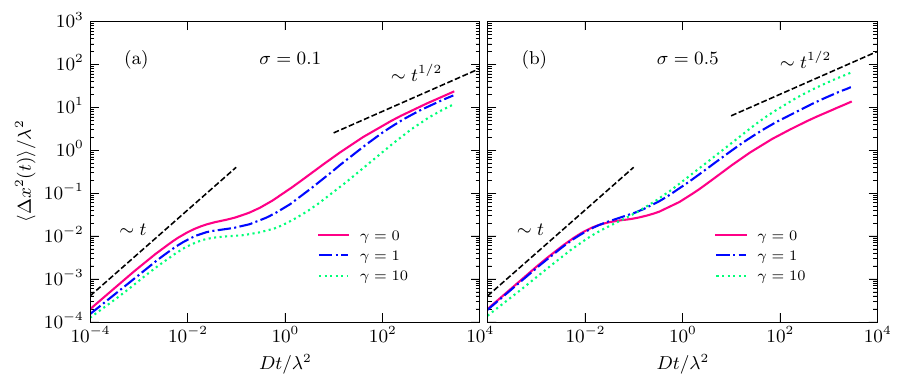}
\caption{Time-dependent mean squared displacements of a tagged particle in a system of sticky hard spheres 
in a sinusoidal periodic potential \eqref{eq:U(x)} at particle number density $\rho=1/2$. 
Hard sphere diameters are $\sigma=0.1$ in (a) and $\sigma=0.5$ in (b). 
The parameter $\gamma$ quantifies the strength of adhesive interaction, see Eq.~\eqref{eq:pair-interaction}. 
Other parameters are $U_0/k_{\rm B}T=6$ and  $L/\lambda=10^3$.}
\label{fig:mean_squared_dispacement}
\end{figure}

Using the BCD algorithm described in Sec.~\ref{sec:algorithm}, we have determined mean squared displacements 
$\langle\Delta x^2(t)\rangle$
of tagged particles in 
equilibrated systems of size $L=10^3\lambda$ with periodic boundary conditions. Results 
are shown in Fig.~\ref{fig:mean_squared_dispacement} for a particle number density $\rho=1/2$ 
and particle diameters (a) $\sigma=0.1$ and (b) $\sigma=0.5$ in the absence of adhesive 
interactions ($\gamma=0$), for strong stickiness ($\gamma=10)$ and an intermediate value $\gamma=1$. 
The ratio of the potential barrier  $U_0$ to the thermal energy is $U_0/k_{\rm B} T=6$.

The results  in Fig.~\ref{fig:mean_squared_dispacement} show a quite complex behavior: while in  
Fig.~\ref{fig:mean_squared_dispacement}(a) the mean squared displacements are smaller for larger $\gamma$ at all 
times, this ordering of the curves is present in Fig.~\ref{fig:mean_squared_dispacement}(b) at short times only. At 
intermediate and large times, the ordering changes in Fig.~\ref{fig:mean_squared_dispacement}(b), i.e.\  mean squared 
displacements are larger for stronger stickiness $\gamma$. Moreover, in (a) there appears a plateau regime when 
$\langle\Delta x^2(t)\rangle/\lambda^2$ is of order 0.01. Such plateau regime is present in (b) only in the absence of 
stickiness ($\gamma=0$).

Based on previous studies in the absence of a periodic potential \cite{Schweers/etal:2023}, we can give a qualitative 
explanation. At short times, a single tagged particle undergoes normal diffusion with the diffusion coefficient $D$. It can be 
part of clusters of different size that move as a whole. The center of mass of an $n$-cluster also diffuses normally at short 
times, but with a reduced diffusion coefficient $D/n$. As clusters become larger with increasing stickiness, the normal 
diffusion slows down with $\gamma$ in both Figs.~\ref{fig:mean_squared_dispacement}(a) and (b). 

At intermediate times, a single tagged particle and a tagged particle in a cluster can partially equilibrate in the potential 
wells. This leads to the plateau in Fig.~\ref{fig:mean_squared_dispacement}(a). A rough estimate of the plateau value
$\langle\Delta x^2\rangle_{\rm pl}$ follows when applying the equipartition theorem for the partial equilibration
of a single tagged particle in a potential well: when $U_0/k_{\rm B}T\gg1$, we can approximate $U(x)$ from 
Eq.~\eqref{eq:U(x)} by the parabolic form $U(x)\simeq (U_0\pi^2/\lambda^2)x^2$, and the equipartition theorem gives  
$\langle\Delta x^2\rangle_{\rm pl}/\lambda^2\simeq k_{\rm B}T/2\pi^2U_0\cong 8.4\times10^{-3}$. 
The plateau is nearly absent for $\gamma=1$ and $\gamma=10$ in Fig.~\ref{fig:mean_squared_dispacement}(b), because 
of 2-cluster formation due to stickiness. The 2-clusters have a size $2\sigma=\lambda$ commensurate with the wavelength 
and can move without surmounting barriers \cite{Antonov/etal:2022a, Antonov/etal:2024}. In 
Fig.~\ref{fig:mean_squared_dispacement}(a) by contrast,
the smallest clusters moving barrier-free have size ten. As the fraction of such 10-clusters is very small, this has no 
relevant impact on the behavior of $\langle\Delta x^2(t)\rangle$. The plateau thus is present for all $\gamma$ in
Fig.~\ref{fig:mean_squared_dispacement}(a).

For longer times, different clusters start to encounter each other and the single-file constraint becomes important. It leads 
to the well-known subdiffusive behavior $\langle\Delta x^2(t)\rangle\sim 2D_{1/2}t^{1/2}$ 
for $t\to\infty$, where $D_{1/2}$ is the subdiffusion coefficient \cite{Schweers/etal:2023}.
In the absence of the periodic potential, the subdiffusion coefficient depends on the number density $\rho$ and the
isothermal compressibility of the system \cite{Kollmann:2003}: the larger the isothermal compressibility the larger 
$D_{1/2}$. With increasing stickiness $\gamma$ and hence clustering of particles, the mean free space between 
neighboring clusters becomes larger.  Accordingly the compressibility increases with $\gamma$ and the subdiffusion 
speeds up. This is reflected in Fig.~\ref{fig:mean_squared_dispacement}(a) by the fact that the curves for
different $\gamma$ approach each other, while they become more separated in 
Fig.~\ref{fig:mean_squared_dispacement}(b). The argument based on our earlier findings in the absence of the periodic 
potential should be valid here, because on large length scales, the effect of the external periodic potential should be 
accounted  for by renormalizing the bare diffusion constant $D$ to an effective one.

The model of sticky hard spheres in a periodic potential with the competing effects discussed above can serve as a useful 
test for advanced theories of diffusion and subdiffusion in single-file systems. It should be possible to develop 
a detailed theory with quantitative predictive power, when determining cluster potentials \cite{Antonov/etal:2024} and 
cluster size distributions \cite{Schweers/etal:2023}.

\section{Conclusions}
\label{sec:conclusions}
We provide an algorithm based on BCD for simulating
single-file systems of hard spheres in external fields with and without
additional particle interactions. The algorithm reduces
the computational demand by multiple orders of magnitude
for large and densely crowded systems.
In the algorithm, updates of particle positions are performed by applying efficient
fragmentation and merging procedures to particle clusters.

The fragmentation of clusters is carried out by sequentially splitting clusters into two subclusters at
those points, where the force difference causing a possible binary splitting is strongest.
Due to this iterative pair splitting procedure, it is not needed to check all $2^{n-1}$ 
possible fragmentations of an $n$-cluster composed of $n$ particles in contact.
Rather it is sufficient to check at most $\mathcal{O}(n^2)$ conditions, and typically even less.

The efficient merging procedure relies on the idea that
mergers of clusters need not to be performed chronologically.
Given the external forces acting on particles in an initial state at the beginning of a time step $\Delta t$, 
the mean velocities of all clusters
are known and one can check, whether neighboring clusters would merge within the interval $\Delta t$. 
Irrespective of when the mergers occur, the respective clusters can be premerged in the initial state by 
placing them at center of mass positions and giving them center of mass velocities. Further premergers 
of allready premerged clusters are carried out until none of the resulting premerged clusters would collide 
with any other premerged cluster in $\Delta t$. 

In fact, our proof of the premerging procedure is at the same time a proof of a general law for completely
inelastic collisions between particles or clusters in one dimension: considering an initial and final state
and constant external forces on particles, the order of the inelastic collisions does not matter, i.e.\ the
same final state is reached for all possible orderings of collisions.

Underlying BCD is the possibility to model particle collisions in overdamped Brownian dynamics 
by completely inelastic collisions due to the absence of inertia. In principle, collisions in overdamped Brownian dynamics
can be treated as elastic, partially inelastic, or completely inelastic.
In inelastic collisions, particles can attach to each other, i.e.\ particle clusters form.

The use of completely inelastic collisions is complementary to the 
treatment by elastic collisions as it is commonly applied in molecular and Brownian 
dynamics simulations. Repetitive collisions between particles in a time step
can be avoided in that case, making the method preferable for simulating highly dense systems.
Apart from that, there are further advantages.
The method allows one to simulate overdamped many-particle dynamics in the limit of
zero noise. 
This was shown recently to be a useful technique to unravel key features of collective particle 
motions in the presence of noise \cite{Antonov/etal:2022a, Antonov/etal:2024}, or, differently speaking,
at finite temperature. We believe that
analyzing overdamped Brownian dynamics in the zero-noise limit can be helpful in many other 
situations where hidden deterministic dynamics dictates striking features in a noisy system.
One can also tackle adhesive interactions, which facilitate cluster formation.
As an example, we have presented simulation results for single-file diffusion
of sticky hard spheres in a sinusoidal potential.
More generally, any additional interaction to the hard sphere repulsion
can be simulated with the BCD method.

\begin{acknowledgments}
Financial support by the Czech Science Foundation (Project
No.\ 23-09074L) and the Deutsche Forschungsgemeinschaft (Project
No.\ 521001072) is gratefully acknowledged.
\end{acknowledgments}


%

\end{document}